\documentclass[final,3p,times,onecolumn]{elsarticle}
\usepackage{lineno,hyperref}

\usepackage{amsmath}
\usepackage{graphicx,epsfig,wrapfig}
\usepackage[caption=false]{subfig}

\usepackage{amsthm,mathrsfs}
\usepackage{makeidx}
\usepackage{amsfonts}
\usepackage{amssymb}
\usepackage{amsmath}
\usepackage{mathtools}
\usepackage{fixmath}
\usepackage{amsbsy}
\usepackage[capitalize]{cleveref}
\usepackage{mathrsfs}
\usepackage{slashed}
\usepackage{bm}
\usepackage{multirow}
\usepackage{booktabs}
\usepackage{adjustbox}

\usepackage{color}
\usepackage[normalem]{ulem}

\definecolor{myGreen}{rgb}{0.2,0.72,0.2}
\renewcommand\sout{\bgroup \color[rgb]{0.55,0.00,0.99} \ULdepth=-.5ex \ULset}

\usepackage{simplewick}

\usepackage{multirow}
\usepackage{braket}

\newcommand{\ta}{\left(}

\newcommand{\tc}{\right)}

\newcommand{\lrDer}[1]{\overset{\leftrightarrow}{#1}\phantom{\,}}
\newcommand{\dilog}[1]{\mathrm{Li}_2\left({#1}\right)}
\newcommand{\LL}{\text{L}_\mu}

\newcommand{\MS}{{\overline{\text{MS}}}}

\newcommand{\e}{\epsilon}

\renewcommand{\[}{\begin{equation}}
\renewcommand{\]}{\end{equation}}

\definecolor{darkgreen}{RGB}{0,120,0}

\allowdisplaybreaks

\modulolinenumbers[5]
\journal{Physics Letters B}
\biboptions{sort&compress}
\begin{document}
\begin{frontmatter}

\title{The gravitational form factors of the electron in quantum electrodynamics}
\author[AFaddr]{Adam Freese}
\ead{afreese@uw.edu}
\author[AMaddr]{Andreas Metz}
\ead{metza@temple.edu}
\author[BPaddr1,BPaddr2]{Barbara Pasquini}
\ead{barbara.pasquini@unipv.it}
\author[SRaddr]{Simone Rodini}
\ead{simone.rodini@polytechnique.edu}

\address[AFaddr]{Department of Physics, University of Washington, Seattle, WA 98195, USA}
\address[AMaddr]{Department of Physics, SERC, Temple University, Philadelphia, PA 19122, USA}
\address[BPaddr1]{Dipartimento di Fisica, Universit\`a degli Studi di Pavia, I-27100 Pavia, Italy}
\address[BPaddr2]{Istituto Nazionale di Fisica Nucleare, Sezione di Pavia, I-27100 Pavia, Italy}
\address[SRaddr]{CPHT, CNRS, Ecole Polytechnique, Institut Polytechnique de Paris, Route de Saclay, 91128 Palaiseau, France}

\begin{abstract}
We calculate the gravitational form factors of the electron at one loop in quantum electrodynamics,
decomposing these into contributions from the electron and photon parts of the energy momentum tensor.
Ultraviolet divergences are removed through renormalization in the $\MS$ scheme.
Infrared divergences are isolated and results are given in both dimensional regularization and photon-mass regularization.
The form factors contain information about the electron's energy and angular momentum structure in QED, as well as its mass radius.
Whenever possible, we compare our results with the existing literature.
\end{abstract}

\date{\today}


\end{frontmatter}

\section{Introduction}
\label{s:introduction}
The matrix elements of the energy momentum tensor (EMT) can be parameterized in terms of the gravitational form factors (GFFs)~\cite{Kobzarev:1962wt, Pagels:1966zza}.
These encode fundamental information about a system, such as the distributions of energy, momentum, angular momentum and internal forces.
The GFFs are therefore of particular interest for hadrons~\cite{Ji:1994av, Ji:1996ek, Polyakov:2002yz, Lorce:2015lna, Lorce:2017wkb, Polyakov:2018zvc, Lorce:2018egm, Freese:2021czn}, where they have been studied in models, lattice QCD and through experimental measurements; see, for instance, Refs.~\cite{Pasquini:2014vua, Hudson:2017oul, Shanahan:2018nnv, Shanahan:2018pib, Pefkou:2021fni, Kumano:2017lhr, Burkert:2018bqq, Kumericki:2019ddg, Dutrieux:2021nlz, Burkert:2021ith}.

Form factors are also important for understanding elementary particles such as the electron.
The best known example is the Pauli form factor of the electron, since its value at vanishing momentum transfer ($t = 0$) gives the anomalous magnetic moment~\cite{Schwinger:1948iu}.
In the present work, we concentrate on the GFFs of the electron as calculated at one loop in quantum electrodynamics (QED).
Partial results can already be found in the literature.
The pioneering one-loop QED calculations of the GFFs have been reported long ago~\cite{Berends:1975ah, Milton:1976jr}.
In contrast to those works, we decompose the GFFs into contributions from the electron and photon parts of the EMT.
Furthermore, we also calculate the form factor appearing in the antisymmetric part of the EMT.
The present study can be considered an extension of our previous work~\cite{Metz:2021lqv} on the electron's GFF $D(t)$, which encodes information about the internal forces of a system~\cite{Polyakov:2002yz, Polyakov:2018zvc}.
Other related studies deal with the QED spin structure~\cite{Brodsky:2000ii, Burkardt:2008ua, Hoyer:2009sg, Liu:2014fxa, Ji:2015sio} and mass structure~\cite{Rodini:2020pis} of the electron.
We also point out that recently the GFFs of the photon were computed at one loop in QED, for both the electron and the photon contributions to the EMT~\cite{Freese:2022ibw}.
Generally, throughout the present work we will refer to, and compare with, previous papers whenever applicable.

The total EMT is ultraviolet finite, but renormalization is required when decomposing the GFFs into contributions from the electron and photon fields.
We present results in the $\MS$ scheme only, being aware that also other renormalization schemes for the EMT are available; see, for instance, Refs.~\cite{Rodini:2020pis, Metz:2020vxd}.
In order to deal with infrared divergences we use dimensional regularization (DR) as well as photon-mass regularization.
In fact, we discuss how the results in one regularization scheme can be transcribed into the other scheme.
At $t = 0$ the form factor $D(t)$ diverges, as was already pointed out previously~\cite{Donoghue:2001qc, Varma:2020crx, Metz:2021lqv}.
Similarly, we find that derivatives of (other) GFFs at $t = 0$ diverge as well.
Those divergences are a consequence of the infinite range of the electromagnetic interaction.
This situation has implications for different radii of the electron which, in principle, could be defined using the GFFs.
For instance, what is normally identified as the mass radius of a particle becomes infinite in the case of the electron.

The paper is organized as follows: In Sec.~\ref{s:definitions} we introduce our main definitions.
Our one-loop analytical results for the GFFs can be found in Sec.~\ref{s:results}, while we use those results to discuss the energy/mass structure and angular momentum structure of the electron in Sec.~\ref{s:quantities}.
In that section, we also address radii of the electron based on the GFFs with a special focus on the electron's mass radius.
In Sec.~\ref{s:summary} we summarize the main findings.

\section{Definitions}
\label{s:definitions}
The EMT is a local operator related to spacetime translation symmetry.
The application of Noether's first theorem to global spacetime translations results in the so-called canonical EMT
as the associated conserved current,
which is in general neither gauge invariant nor symmetric.
This is the case in QED, where the symmetry properties of the Lagrange density give the following (conserved) canonical EMT:
\begin{equation}
T^{\mu\nu}_{\text{can.}} = \bar{\psi}\gamma^\mu\frac{i\lrDer{\partial}^\nu}{2}\psi - F^{\mu\alpha}F^{\nu}_{\ \ \alpha}+\frac{g^{\mu\nu}}{4}F^2 - F^{\mu\rho}\partial_\rho A^\nu .
\end{equation}
This operator can be made gauge-invariant following the Belinfante-Rosenfeld procedure~\cite{Belinfante1,Belinfante2,Rosenfeld} of adding a divergence term according to
\begin{equation}
T^{\mu\nu} = T^{\mu\nu}_{\text{can.}} +  \partial_\rho \Phi^{\rho\mu\nu} ,
\end{equation}
where the so-called superpotential reads
\begin{equation}
 \Phi^{\rho\mu\nu} = F^{\mu\rho}A^\nu.
\end{equation}
Note that the superpotential can be interpreted as the spin operator for the photon field, which is well known to be not gauge-invariant.
As a result, we obtain
\begin{align}
T^{\mu\nu} & = T_e^{\mu\nu}+T_\gamma^{\mu\nu}, \;\, \textrm{with} \\
 T_e^{\mu\nu} & =\bar{\psi}\gamma^\mu\frac{i\lrDer{D}^\nu}{2}\psi, \quad\quad
 T_\gamma^{\mu\nu}=- F^{\mu\alpha}F^{\nu}_{\ \ \alpha}+\frac{g^{\mu\nu}}{4}F^2,
\end{align}
where we neglected the contributions from the equation of motions (EOMs) and the gauge-fixing term, which vanish when evaluated in matrix elements for physical states.
We note that the EMT $T_\gamma^{\mu\nu}$ in the gauge sector is now symmetric, while the electron contribution $T_e^{\mu\nu}$ has still a (gauge-invariant) antisymmetric part, associated with the electron spin contribution.
We stress that no gauge-invariant definition of a photon spin contribution is possible;
such a term would need to be the divergence of a superpotential $\partial_\rho \Lambda^{\rho\mu\nu}$ with three Lorentz indices, which cannot be constructed using only the field strength tensor $F_{\mu\nu}$.
In the nomenclature of Ref.~\cite{Leader:2013jra}, $T^{\mu\nu}$ is called the gauge-invariant kinetic EMT.

In this work we focus on the matrix elements of the EMT between single-electron states, while we refer to~\cite{Berends:1975ah, Milton:1977je, Freese:2022ibw} for the results between single-photon states.
We use the parametrization~\cite{Kobzarev:1962wt,Pagels:1966zza,Ji:1996ek}
\begin{equation}
\braket{p',s' |T_i^{\mu\nu}| p,s} = \bar{u}(p',s') \Bigg( A_i \frac{P^\mu P^\nu}{m} + (A_i+B_i)\frac{iP^{\{\mu}\sigma^{\nu\}\Delta}}{4m}+C_i \frac{\gamma^{[\mu}P^{\nu]}}{2}  + D_i \frac{\Delta^\mu \Delta^\nu - g^{\mu\nu}\Delta^2}{4m} + \bar{C}_i mg^{\mu\nu} \Bigg) u(p,s),
\label{parametrization_T}
\end{equation}
where $p' = P+\frac{\Delta}{2}$, $p=P-\frac{\Delta}{2}$, $i=e,\gamma$, and $u(p,s)$ is the electron spinor with $\bar{u}(p,s) u(p,s) = 2m$.
We also used shorthand notations for symmetrization ($a^{\{\mu}b^{\nu\}} = a^\mu b^\nu + a^\nu b^\mu$) and antisymmetrization ($a^{[\mu}b^{\nu]} = a^\mu b^\nu - a^\nu b^\mu$) of indices, as well as $a^\Delta = a^\mu \Delta_\mu$.
In Eq.~\eqref{parametrization_T} the functions $A_i,B_i,D_i,C_i,\bar{C}_i$, known as the gravitational form factors (GFFs), are functions of the invariant $t=\Delta^2$ and the renormalization scale $\mu$.
Due to conservation of the EMT, the sum of the electron and photon contributions to any GFF is renormalization-scale invariant, and
$\bar{C}(t)=\sum_i \bar{C}_i(t, \mu)=0$.
The conservation of the EMT also leads to the constraints
\begin{align}
A_e(0, \mu) + A_\gamma(0, \mu) &= 1,
\label{e:mom_cons}\\
\sum_i \frac{1}{2} \big( A_i(0, \mu) + B_i(0, \mu) \big) & = \sum_i J_i(\mu) = \frac{1}{2},
\label{e:angmom_cons}
\end{align}
which, respectively, express the conservation of momentum and of the total angular momentum $J = J_e(\mu) + J_\gamma(\mu)$.
(Note that $\frac{1}{2} \big( A(t)+B(t)\big)$ is also called the angular-momentum form factor; see, for instance, Ref.~\cite{Polyakov:2002yz}.)
The relations~\eqref{e:mom_cons} and~\eqref{e:angmom_cons} readily imply
\begin{equation}
B_e(0, \mu) + B_\gamma(0, \mu)=0,
\label{e:B_constraint}
\end{equation}
which is often referred to as the vanishing of the anomalous gravitomagnetic moment~\cite{Pagels:1966zza, Ji:1996ek, Teryaev:1999su, Brodsky:2000ii, Silenko:2006er, Teryaev:2016edw, Lowdon:2017idv}.
The form factors $D_i$ (and $\bar{C}_i$), which may be related to pressure and shear distributions~\cite{Polyakov:2002yz, Polyakov:2018zvc}, can in principle take on any values for $t = 0$.
The same applies for the form factor $C$, which is associated with the antisymmetric part of the EMT and therefore has a vanishing photon contribution.
Using the QED equation of motion, one finds that $C$ is equivalent to the axial form factor~\cite{Leader:2013jra,Lorce:2017wkb}, that is,
\begin{equation}
C(t, \mu) = C_e(t, \mu) = G_A(t, \mu).
\end{equation}

The renormalization of the EMT is well understood in the literature; see, for instance, Ref.~\cite{Rodini:2020pis} for QED and Refs.~\cite{Hatta:2018sqd, Tanaka:2018nae, Metz:2020vxd} in the case of QCD.
All the counterterms relevant for our one-loop calculation in the $\MS$ scheme can be found in Ref.~\cite{Rodini:2020pis}.
The antisymmetric  part of the EMT does not require any additional renormalization (beyond the renormalization of the parameters of the Lagrange density). Since there is no other rank-2 tensor operator with the appropriate mass dimensions, there can be no mixing between the antisymmetric part of the EMT with any other operator. In principle the antisymmetric part of the EMT could require a multiplicative renormalization, but this is not the case in QED at one-loop order.
Instead of renormalizing the full EMT right away, one may exploit the fact that the total EMT can be decomposed into a trace and traceless part, both of which are separately scale-invariant.
One can then further decompose those two terms into electron and photon contributions and renormalize them.
We have gone through this procedure as well, finding complete agreement for the results of all the GFFs.

\section{Results for the gravitational form factors}
\label{s:results}
\begin{figure}[t]
    \centering
    \includegraphics[width = 0.75\textwidth]{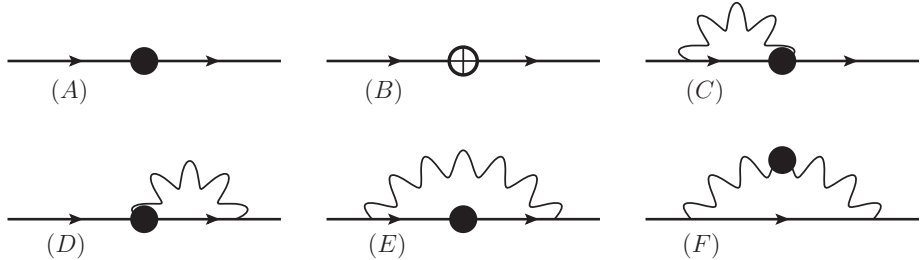}
    \caption{Relevant diagrams for insertion of the EMT operator in the $e\to e$ amplitude at one-loop. The black dot represents the EMT insertion, the white crossed dot represents the EMT insertion accompanied by the wave function renormalization for the electron.}
    \label{diagrams}
\end{figure}
The aim of this section is to present the QED results of one-loop calculation for the GFFs.
The relevant diagrams are shown in Fig.~\ref{diagrams}.
The total electron contribution does not depend on the subtraction scheme used for the Lagrangian counterterms.
Therefore we chose to work in the on-shell scheme, in which there is no contribution when the loop is confined to a single electron leg.
Moreover, the individual diagrams for the electron contribution are not gauge invariant.
We will present only their sum, for which gauge invariance is restored.

In our one-loop calculation only the form factors $A_i$ and $\bar{C}_i$ exhibit ultraviolet (UV) divergences, which are removed by operator renormalization.
At finite momentum transfer, the form factors $A_e$ and $C_e$ also show a (standard) infrared (IR) divergence.
The IR divergences are expected to be canceled in physical cross sections by similar IR divergences from soft final-state radiation.
In order to regulate them we use both DR and photon-mass regularization.
To present the results in DR we introduce $L_\mu = \log (\bar{\mu}^2 / m^2)$, where $\bar{\mu}^2 = \mu^2 4\pi e^{-\gamma_E}$, with the DR scale $\mu$ and the Euler constant $\gamma_E$.

The results for the GFFs are obtained by inserting the EMT operator into a Green's function via the LSZ reduction theorem.
After isolating the various independent structures, we obtain
\begin{align}
A_e &=1-\frac{2\alpha}{3\pi}\log \frac{\mu^2}{m^2} +\frac{\alpha}{4\pi}\left(\frac{1}{\e}+\LL\right)\left[\left(\nu +\frac{1}{\nu }\right) \log \left(\frac{1+\nu}{1-\nu}\right)-2\right] +\frac{\alpha}{8\pi\nu} \left(1+\nu ^2\right) \Bigg\{
2 \dilog{\frac{1+\nu }{2}} -2
   \dilog{\frac{1-\nu}{2}} \notag \\
& + \log
   \left(\frac{4 \nu }{1-\nu}\right) \log (\nu(1-\nu))
  - \log
   \left(\frac{4 \nu }{1+\nu}\right) \log (\nu(1+\nu))
   +\frac{1}{3} \frac{16\nu ^2+18}{1+\nu^2}\log \left(\frac{1+\nu}{1-\nu}\right) -\frac{176 \nu}{9(1+\nu^2)}
\Bigg\},\\
B_e &= \frac{\alpha}{6\pi\nu}\left(1-\nu ^2\right) \log \left(\frac{1+\nu}{1-\nu }\right), \\
C_e &= 1 +\frac{\alpha}{4\pi}\left(\frac{1}{\e}+\LL\right)\left[\left(\nu +\frac{1}{\nu }\right) \log \left(\frac{1+\nu}{1-\nu}\right)-2\right] +\frac{\alpha}{8\pi\nu}\left(1+\nu ^2\right)\Bigg\{2  \dilog{\frac{1+\nu}{2}}-2
   \dilog{\frac{1-\nu}{2}}\notag \\
   & +
  \log \left(\frac{4 \nu }{1-\nu}\right) \log (\nu(1-\nu))-\log \left(\frac{4 \nu }{1+\nu }\right) \log (\nu  (1+\nu))+2\frac{1+2 \nu ^2}{1+\nu^2} \log \left(\frac{1+\nu}{1-\nu }\right)
\Bigg\}, \\
D_e &=\frac{5\alpha}{12\pi}\frac{\left(1-\nu ^2\right) }{\nu ^3}\left(2\nu -\log\ta\frac{1+\nu}{1-\nu}\tc\right), \\
\bar{C}_e &= \frac{\alpha}{6\pi} \log \frac{\mu^2}{m^2} -\frac{\alpha}{18\pi},\\
A_\gamma &= \frac{2\alpha}{3\pi} \log \frac{\mu^2}{m^2} + \frac{\alpha}{12\pi}\Bigg\{ \frac{34}{3} -10\nu^2(1-\nu^2) - \left(15 \nu ^4-22 \nu
   ^2+15\right)  \nu ^2\log \left(4\eta \right) -3 \nu (1-\nu^2) \left(5 \nu^4 - 4\nu^2 + 3\right) \mathcal{F}(\nu)\Bigg\}, \\
B_\gamma &= \frac{\alpha}{12\pi}(1-\nu^2)\Bigg\{10\nu^2 + 3\nu\left(5 \nu ^4-6 \nu ^2+1\right) \mathcal{F}(\nu)-\left(15 \nu^2-13 \right)\nu^2 \log \left(4\eta\right)-4\Bigg\} ,\\
C_\gamma &= 0, \\
D_\gamma &= \frac{\alpha}{12\pi \nu}(1-\nu^2)^3\Bigg\{ \frac{2 \nu }{\left(1-\nu ^2\right)^2}+\frac{ \nu  \left(5-3 \nu ^2\right) \log \left(4\eta\right)}{\left(1-\nu ^2\right)^2}+3\mathcal{F}(\nu)\Bigg\}, \\
\bar{C}_\gamma &= -\frac{\alpha}{6\pi} \log \frac{\mu^2}{m^2} +\frac{\alpha}{18\pi} ,
\end{align}
where we defined
\begin{equation}
\eta = \frac{-t}{4m^2}, \quad \nu = \sqrt{\frac{\eta}{1+\eta}}
\end{equation}
and
\begin{align}
\mathcal{F}(\nu) &= \frac{\pi^2}{3} + \frac{1}{4} \log^2\left(\frac{1-\nu}{1+\nu}\right)+\dilog{\frac{1-\nu}{1+\nu}}, \quad \text{ with }\
  \dilog{z}   =   -   \int_0^z \mathrm{d}t  \,   \frac{\log(1-t)}{t}.
\end{align}
In the above results we used DR to regulate the IR divergences present in $A_e$ and $C_e$.
The corresponding results with a nonzero photon-mass regulator ($m_\gamma$) are obtained by the replacement
\begin{equation}
\frac{1}{\e} + \LL \to \log \frac{m_\gamma^2}{m^2}.
\end{equation}
When summing the electron and photon contributions of the GFFs $A$, $B$ and $D$, we recover the results that were already presented in Refs.~\cite{Berends:1975ah, Milton:1976jr}.
More precisely, for the full $t$-dependence of those form factors we found agreement in a careful numerical comparison,
while we found exact analytic agreement
for the small and large $t$ limits (to be discussed below) and for the IR divergent part present in the GFF $A$.
The results for $D_i$ and $\bar{C_i}$ were already discussed in Ref.~\cite{Metz:2021lqv}, though the exact analytic forms for $D_i$ presented here are new.
Interestingly, the GFFs $\bar{C}_i$ are constant.
To the best of our knowledge, the GFF $C$ has not been computed before.

The $\log (\mu^2 / m^2)$ term in the $A_i$ and $\bar{C}_i$ is a leftover of operator renormalization and cancels when summing the respective electron and photon contributions.
We also point out that the IR divergence cancels in the linear combination $(A_e + B_e - C_e)/2$, which can be considered the form factor associated with the so-called kinetic orbital angular momentum (OAM) $L_e^{\rm kin}$ appearing in the spin sum rule by Ji~\cite{Ji:1996ek}.
The form factor $D(t)$ diverges at $t = 0$, with the divergence arising from the photon contribution $D_{\gamma}(t)$.
This divergence was discussed previously~\cite{Donoghue:2001qc, Varma:2020crx, Metz:2021lqv} and is a consequence of the infinite range of the electromagnetic interaction.

It is interesting to isolate the behavior of the GFFs both in the limit of vanishing momentum transfer and of large momentum transfer.
We find in the small-$t$ limit
\begin{align}
A_e(t) &= 1 - \frac{\alpha}{18\pi}\left( 17 + 12 \log \frac{\mu^2}{m^2} \right)+\frac{\alpha}{18\pi} \left[27+12\left(\frac{1}{\epsilon} +L_\mu\right)\right]\eta + {\cal O}(\eta^2) ,
\label{e:Ae_low_t}
\\
B_e(t) &= \frac{\alpha}{3\pi} - \frac{2\alpha}{9\pi}\eta + {\cal O}(\eta^2) ,
\\
C_e(t) &= 1+\frac{\alpha}{2\pi}+\frac{\alpha}{6\pi}\left[5+4\left(\frac{1}{\epsilon} +L_\mu\right)\right]\eta + {\cal O}(\eta^2) ,
\\
D_e(t) &= -\frac{5\alpha}{18\pi}+\frac{\alpha}{9\pi}\eta + {\cal O}(\eta^2) ,
\\
A_\gamma(t) &= \frac{\alpha}{18\pi} \left( 17 + 12 \log \frac{\mu^2}{m^2} \right)-\frac{3\alpha \pi}{8}\sqrt{\eta} + \frac{2\alpha}{3\pi}(1-3\log(4\eta))\eta + {\cal O}(\eta^{3/2}) ,
\label{e:Ag_low_t}
\\
B_\gamma(t) &= -\frac{\alpha}{3\pi}+\frac{\alpha \pi}{8}\sqrt{\eta}+ \frac{2\alpha}{3\pi}(1+2\log(4\eta))\eta + {\cal O}(\eta^{3/2}) ,
\\
D_\gamma(t) &= \frac{\alpha\pi}{8\sqrt{\eta}}-\frac{\alpha}{3\pi}(1-2\log(4\eta))-\frac{5\alpha\pi}{16}\sqrt{\eta} + \frac{2\alpha}{9\pi}(5-6\log(4\eta))\eta + {\cal O}(\eta^{3/2}).
\label{e:Dg_low_t}
\end{align}
Note that our results for the $A_i$ and $B_i$ are compatible with the constraints in Eq.~\eqref{e:mom_cons} and Eq.~\eqref{e:B_constraint}, respectively.
Furthermore, we agree with the result for the gravitomagnetic moment $B_e(0,\mu) = - B_\gamma(0, \mu) = \alpha / 3\pi$ reported previously~\cite{Brodsky:2000ii}.
In the large-$t$ limit, the GFFs behave as
\begin{align}
A_e(t) &= -\frac{\alpha}{4\pi}\log^2(\eta) + {\cal O}(\log\eta) ,
\\
B_e(t) &= \frac{\alpha}{\pi}\frac{\log(4\eta)}{6\eta} + {\cal O} \left( \frac{\log\eta}{\eta^2} \right) ,
\\
C_e(t) &= -\frac{\alpha}{4\pi}\log^2(\eta) + {\cal O}(\log\eta) ,
\\
D_e(t) &= \frac{\alpha}{\pi}\frac{10-5\log(4\eta)}{12\eta} + {\cal O} \left( \frac{\log\eta}{\eta^2} \right) ,
\\
A_\gamma(t) &= -\frac{2\alpha}{3\pi}\log(4\eta) + {\cal O}(\eta^0) ,
\\
B_\gamma(t) &= \frac{\alpha}{\pi}\frac{3-\log(4\eta)}{6\eta} + {\cal O} \left( \frac{\log\eta}{\eta^2} \right) ,
\\
D_\gamma(t) &= \frac{\alpha}{\pi}\frac{1+\log(4\eta)}{6\eta} + {\cal O} \left( \frac{\log\eta}{\eta^2} \right) .
\end{align}
Notice that the $\eta$-dependence of the IR pole in $A_e$ and $C_e$ is always suppressed compared to the leading behavior of the form factors.
For the $D_i$ such expansions were already given in Ref.~\cite{Metz:2021lqv}.
We repeat that, after summing over the electron and photon contributions, we find agreement with the expansions of the GFFs $A$, $B$, and $D$ presented in Refs.~\cite{Berends:1975ah, Milton:1976jr}.
We note that these are strictly perturbative results and agree with the perturbative QCD results for a quark target (modulo a color factor). In Refs.~\cite{Tong:2021ctu,Tong:2022zax,Tanaka:2018wea}, gravitational form factors have been calculated in perturbative QCD for hadrons at large momentum transfer. However, for a bound state, the large-$t$ behavior is generally different from pointlike particles.

\section{Quantities of special interest}
\label{s:quantities}
The GFFs contain a wealth of information about a system.
Here we will briefly discuss the resulting energy/mass structure and angular momentum structure of the electron in QED, as well as different electron radii.
We note that mechanical properties of the electron, which follow from the GFFs $D_i$ and $\bar{C}_i$~\cite{Polyakov:2002yz, Polyakov:2018zvc}, were already discussed previously~\cite{Metz:2021lqv}.

The expectation value of $T^{00}$ for a physical state furnishes an energy density at fixed instant-form time.
Integrating this over all space gives the total energy,
which for a zero-momentum plane wave state is equal to the mass.
Accordingly, EMT matrix elements at $P=(m;0,0,0)$ and $\Delta=0$ provide a mass decomposition~\cite{Ji:1994av},
which is fully determined by $A_i(0, \mu)$ and $\bar{C}_i(0, \mu)$.
The aforementioned constraints on those GFFs then imply that two numbers only fix the EMT of a spin-$\frac{1}{2}$ particle in the forward limit~\cite{Ji:1996ek, Lorce:2017xzd}.
Different mass sum rules have been proposed in the literature; see, for instance, Refs.~\cite{Ji:1994av, Ji:1995sv, Hatta:2018sqd, Tanaka:2018nae, Lorce:2017xzd, Rodini:2020pis, Metz:2020vxd, Ji:2021qgo, Lorce:2021xku}.
Here we refrain from listing the corresponding one-loop QED results for the electron.
We just mention that the one-loop $\MS$-results for a 3-term mass sum rule were already provided in~\cite{Rodini:2020pis}, while expressions for any other mass sum rule can be readily obtained from our results for the $A_i(0, \mu)$ and $\bar{C}_i(0, \mu)$.

We now turn our attention to the angular momentum structure of the electron; see also Refs.~\cite{Brodsky:2000ii, Burkardt:2008ua, Hoyer:2009sg, Liu:2014fxa, Ji:2015sio}.
For the electron and photon contributions to the total angular momentum (spin) of the (physical) electron we find
\begin{equation}
J_e(\mu) = \frac{1}{2}-\frac{\alpha}{3\pi}\left(\frac{11}{12}+ \log \frac{\mu^2}{m^2} \right),\quad
J_\gamma(\mu) = \frac{\alpha}{3\pi}\left(\frac{11}{12}+ \log \frac{\mu^2}{m^2} \right),
\end{equation}
which agrees with the expressions obtained in Ref.~\cite{Ji:2015sio} (after applying $\MS$-subtraction to the results of that paper).
The form factor $C_e$ provides the electron spin contribution to the total angular momentum of the (physical) electron according to
\begin{equation}
S_e=\frac{1}{2}C_e(0)=\frac{1}{2}\left(1+\frac{\alpha}{2\pi}\right),
\end{equation}
which again agrees with the result reported in~\cite{Ji:2015sio}.
Note that the one-loop expression of $C_e$ is scale-invariant.
Hence one obtains the electron contribution to the kinetic OAM~\cite{Ji:1996ek} as
\begin{equation}
L_e^{\rm kin}(\mu) = J_e(\mu) - S_e =
-\frac{\alpha}{3\pi}\left(\frac{5}{3} + \log \frac{\mu^2}{m^2} \right).\label{eq:OAM}
\end{equation}
As shown in Ref.~\cite{Ji:2015sio}, this result coincides with the canonical OAM of the Jaffe-Manohar (JM) sum rule~\cite{Jaffe:1989jz} at order ${\cal O}(\alpha)$.
The difference between the JM and kinetic OAM is known as potential angular momentum~\cite{Wakamatsu:2010qj}. From the calculations of the EMT, the potential angular momentum can be derived from diagrams (C) and (D) in Fig.~\ref{diagrams}. These diagrams give a contribution to the form factor $J_e$ that is canceled by an identical contribution to $S_e$, making the final contribution to the OAM in Eq.~\eqref{eq:OAM} equal to zero.
We therefore confirm the result of Ref.~\cite{Ji:2015sio} that  the potential angular momentum of the electron is vanishing at one-loop order.

The results for the GFFs, a priori, also allow one to compute radii of a particle.
Of particular interest is the mass radius, which is related to the form factor of $T^{00}$ and can be calculated according to
\begin{equation}
\langle r^2 \rangle_{\rm m} =   \left[6 \frac{d A (t)}{dt} -  \frac{3}{2}\frac{D(t)}{m^2} \right]_{t = 0},
\label{e:mass_radius}
\end{equation}
(see, for instance, Refs.~\cite{Kharzeev:2021qkd, Ji:2021mtz}).
The mass radius of the electron is ill-defined;
the small-$t$ behavior of both $A'(t)$ and $D(t)$ is singular at $t=0$---as can be verified from
Eqs.~\eqref{e:Ag_low_t} and \eqref{e:Dg_low_t}---and the singularities do not cancel.
(The aforementioned divergence in $D(0)$ also prevents one from defining a meaningful mechanical radius of the electron~\cite{Metz:2021lqv}.)
Furthermore, there is no linear combination of $A'(t)$ and $D(t)$ that is finite at $t=0$,
since both functions exhibit both $1/\sqrt{-t}$ and $\log(|t|)$ behavior at small $t$ that cannot be simultaneously canceled,
making it impossible to define an alternative mass radius which is finite.
Additionally, the result for $A_e$ (see Eq.~\eqref{e:Ae_low_t}) implies that also a ``standard" IR divergence is present in the mass radius.
All of these same complications arise when trying to define a radius through the form factors associated with the trace of the EMT or the kinetic OAM.
The radius of the (spin) form factor $C$ is not plagued by the singularity at $t = 0$, but still suffers from an IR divergence.
In passing we repeat that the peculiar singular behavior at $t = 0$ results from the infinite range of the electromagnetic interaction.
In principle, such (infinite) contributions could also be present for charged hadrons in quantities like the mass radius.
(A related discussion in the case of $D(0)$ can be found in Refs.~\cite{Varma:2020crx, Metz:2021lqv}.)

\section{Summary}
\label{s:summary}
We have presented a complete calculation of the separate electron and photon contributions to the GFFs of the electron at one-loop accuracy in QED,
thus going beyond the pioneering papers~\cite{Berends:1975ah, Milton:1976jr} in which the full GFFs were presented.
To the best of our knowledge, we calculated the form factor associated with the antisymmetric part of the EMT for the first time.
At one loop, some of the GFFs exhibit UV divergences which we addressed using operator renormalization in the $\MS$ scheme.
We regulated IR divergences by employing both dimensional regularization and photon-mass regularization.
Among other things, the GFFs contain information on the electron's mass structure, angular momentum structure, and mass radius.
We confirmed the interesting result that, at one loop in QED, the electron kinetic OAM and the canonical OAM are identical~\cite{Ji:2015sio}.
Furthermore, we find that a meaningful mass radius for the electron cannot be defined.
One obstacle is a (standard) IR divergence,
but more importantly, the infinite range of the electromagnetic interaction gives rise to irremovable singularities at $t = 0$.
We close by repeating that the one-loop results for the GFFs of the electron are fundamental QED results.

\section*{Acknowledgement}
This work of A.F.~was supported by the U.S.~Department of Energy Office of Science, Office of Nuclear Physics under Award Number DE-FG02-97ER-41014.
The work of A.M.~was supported by the National Science Foundation under the Grant No.~PHY-2110472, and by the U.S. Department of Energy, Office of Science, Office of Nuclear Physics, within the framework of the TMD Topical Collaboration.
The work of B.P. is part of a project that has received funding from the European Union’s Horizon 2020 research and innovation programme under grant agreement STRONG – 2020 - No~824093.




\end{document}